# A comparative study of the high-pressure structural stability of zirconolite materials for nuclear waste immobilisation

Daniel Errandonea[1,*], Robin Turnbull[1], Josu Sánchez-Martín[1], Robert Oliva[2], Alfonso Muñoz[3], Silvana Radescu[3], Andres Mujica[3], Lewis Blackburn[4], Neil C. Hyatt[5,6], Catalin Popescu[7], Jordi Ibáñez-Insa[2]

[1]Departamento de Física Aplicada-ICMUV, MALTA-Consolider Team, Universidad de Valencia, Dr. Moliner 50, Burjassot, 46100 Valencia, Spain

[2]Geosciences Barcelona (GEO3BCN), MALTA-Consolider Team, Spanish Council for Scientific Research (CSIC), Lluís Solé i Sabarís s/n, 08028 Barcelona, Spain

[3]Departamento de Física, MALTA-Consolider Team, Instituto de Materiales y Nanotecnología, Universidad de La Laguna, San Cristóbal de La Laguna, E-38200 Tenerife, Spain

[4]Department of Materials Science and Engineering, The University of Sheffield, United Kingdom

[5]Building 329, Thomson Avenue, Harwell Campus, Didcot, OX11 0GD, United Kingdom

[6]School of Earth Sciences, The University of Bristol, Bristol, BS8 1RL, United Kingdom

[7]CELLS-ALBA Synchrotron Light Facility, Cerdanyola del Vallès, 08290 Barcelona, Spain

*Corresponding author: E-mail: daniel.errandonea@uv.es

**Abstract:** We present a comparative study of the high-pressure behaviours of the nuclear waste immobilisation materials zirconolite-2M, -4M, -3O, and -3T. The materials are studied under high-pressure conditions using synchrotron powder X-ray diffraction. For zirconolite-2M we also performed density-functional theory calculations. A new triclinic crystal structure (space group $P\bar{1}$), instead of the previously assigned monoclinic structure (space group $C2/c$) is proposed for zirconolite-2M. We named the triclinic structure as zirconolite-2TR. We also found that zirconolite-2TR undergoes a phase transition at 14.7 GPa to a monoclinic structure described by space group $C2/c$, which is different than the high-pressure structure previously proposed in the literature. These results are discussed in comparison with previous studies on zirconolite-2M and the related compound calzirtite. For the other three zirconolite structures (4M, 3O, and 3T) this is the first high-pressure study, and we find no evidence for pressure induced phase transitions in any of them. The linear compressibility of the studied compounds, as well as a room-temperature pressure-volume equation of state, are also presented and discussed.

## 1. Introduction

zirconolite, prototypically $CaZrTi_2O_7$, is one of several titanate compounds that can be used in synthetic ceramic waste-forms designed to immobilise actinides and fission products arising from nuclear fuel reprocessing [1-5]. As a mineral species, zirconolites have been shown to retain their original U and Th inventories for hundreds of millions of years, much longer than the safety assessment period for a deep geological repository [6-8]. Trivalent and tetravalent actinides and lanthanides are capable of isomorphic substitution at the Ca and/or Zr sites, given compatible ionic radii, with potential charge compensation by co-substitution, including at the smaller Ti sites, as necessary [6-8]. Zirconolite ceramics are a candidate waste-form for immobilisation of any civil separated plutonium not reused as fuel, due to the chemical flexibility of the zirconolite structure, high capacity for actinide incorporation, and chemical and radiation stability over geological timescales [6-8].

The crystal structure of zirconolite ($CaZrTi_2O_7$) is related to the cubic pyrochlore structure with the chemical formula $A_2B_2O_7$ (where A represents an 8-fold coordinated cation, such as Ca, Sr, Pb, Bi, etc. and B represents 6-fold coordinated cations such as Ti, Sn, Zr, Hf, etc.). Several zirconolite polytype structures exist, namely: 2M, 3T, 3O, and 6T. On the other hand, the zirconolite-4M structure is a 1:1 ordered intergrowth of zirconolite and pyrochlore unit cells parallel to the (001) planes [4]. The aristotype structure is that of zirconolite-2M, for which the crystal structure has a monoclinic lattice in the *C*2/*c* space group [9]. Zirconolite-2M has been described as a layered structure consisting of alternating $CaO_8$ and $ZrO_7$ polyhedra in rows parallel to (110) planes. The titanium atoms form a hexagonal tungsten bronze (HTB) motif [10]. In the polytype nomenclature, e.g. zirconolite-2M, the number denotes the number of repeating HTB layers in the unit-cell and the letter denotes the crystal system as monoclinic (M), trigonal (T), or orthorhombic (O). Zirconolite-T is described by space group $P3_121$ [11], zirconolite-3O is described by space group *Cmce* [12], while the crystal structure of zirconolite 6T has not yet been solved. zirconoline-4M is described by space group *C*2/*c* [11] as in the case of zirconolite-2M. Whereas zirconolite-2M, -3T and -3O, are polytype structures, related by the stacking arrangement of HTB layers, zirconolite-4M is a distinct intergrowth structure.

Understanding the behaviour of ceramic waste-forms, such as zirconolite, under high pressure conditions is important and useful from a number of perspectives. Compressibility data provide a direct measurement of the bulk modulus and elastic constants, which are essential for determining mechanical properties relevant to structural integrity and for validating of computational models of crystal structure. Pressure-induced structural phase transitions and structural disorder can provide insight into alternative atomic arrangements, with potentially enhanced chemical or physical properties, which may be stabilised by appropriate "chemical pressure" induced by engineering of the crystal chemistry. Pressure-induced crystalline-to-amorphous phase transitions may also be a useful proxy for the progressive loss of long range structural order resulting from the long-term accumulation of alpha recoil damage in ceramic waste-forms, where the rate of damage exceeds the rate of recovery. Indeed, pressure is known to have a significant influence on the crystal structures and physical properties of solids by inducing structural phase transitions [13], inducing structural disorder [14], inducing crystalline-to-amorphous phase transitions [15], or inducing chemical decomposition [16].

To date, several high-pressure studies have been performed on perovskite-type $CaTiO_3$ [17] and $CaZrO_3$ [18]. Phase transitions have been reported at pressures around 30 GPa [17, 18] and the pressure-volume equations of state have been determined, showing that these perovskites have bulk moduli in the range 170-200 GPa [17, 18]. In contrast, only a few studies have been performed to investigate the structural stability and high-pressure properties of the quaternary Ca-Zr-Ti oxides. These studies have focused on calzirtite $Ca_2Zr_5Ti_2O_{16}$ [19] and zirconolite $CaZrTi_2O_7$ [20]. Calzirtite has been studied by powder X-ray diffraction (XRD) under high pressure (HP) up to 30 GPa. Evidence of a structural phase transition was found at 12–13 GPa and the crystal structure of the HP phase was identified [19]. Zirconolite-2M has been studied by the same method up to 62 GPa and reported to undergo a phase transition from *C*2/*c* to *P*2$_1$/*m* above 15.6 GPa. However, only the space group and unit cell parameters were proposed for the HP phase [20]. Compression of zirconolite-2M above 56 GPa was reported to induce a transition to a disordered meta-stable phase with a cottunite-related structure, which could be recovered to ambient pressure. Such amorphization has previously been reported in pyrochlore-related systems under pressure [6]. All these results suggest that further studies are needed to fully understand the HP behavior of Ca-Zr-Ti oxides.

In this work, we have investigated the HP behavior of four different calcium zirconium titanate compounds: zirconolite-2M ($CaZrTi_2O_7$), zirconolite-4M ($Ca_{0.75}Zr_{0.75}Dy_{0.50}Ti_2O_7$), zirconolite-3O ($Ca_{0.20}Nd_{0.80}ZrTi_{1.20}Fe_{0.80}O_7$), and zirconolite-3T ($Ca_{0.65}Ce_{0.35}ZrTi_{1.30}Fe_{0.70}O_7$). Here, all four compounds have been studied at ambient temperature under HP up to 30 GPa using synchrotron-based powder XRD. We found that zirconolite-4M, 3O, and 3T do not undergo any structural phase transition in the pressure range considered. In contrast, in zirconolite-2M a phase transition has been found at 14.7 GPa. In this compound, by combining Rietveld refinements of XRD patterns and density functional theory calculations, we have found that the crystal structure at ambient pressure is most likely triclinic rather than monoclinic as previously reported in the literature. In particular, we have found that the monoclinic structure [10] leads to dynamically unstable imaginary phonon branches. We have also determined the crystal structure of the HP phase, which is monoclinic. From XRD experiments we have also determined the linear compressibility of the axes of each of the four structures as well as a pressure-volume equation of state, which allows us to compare the compression behaviour of the different zirconolite phases. The reported study will help to understand the crystal chemistry and physical properties of zirconolite and related compounds and contribute to the design of nuclear-waste storage materials. In special, they are of particular relevance for the construction of deep geological repositories for storing hazardous or radioactive waste within a stable geologic environment.

**2. Methods**

*2.1 Experiments*

A solid-state synthesis method was used for the preparation of zirconolite-2M, zirconolite-4M, zirconolite-3O, and zirconolite-3T samples from the constituent oxides [21]. High purity oxide precursors were used in the synthesis. The precursors were batched according to the desired compositions of each zirconolite structure by roller milling in a $ZrO_2$ medium for 24 h, with acetone added as a milling agent. After drying, the ground precursor material was compacted under 100 MPa uniaxial pressure. The resulting pellets were placed in a furnace and sintered in air at 1350 °C, for a dwell time of 20 h. The nominal compositions of the samples are those described in the introduction (see previous paragraph).

Powder XRD experiments under HP were carried out at the BL04-MSPD beamline at the ALBA-CELLS synchrotron [22]. The experiments were performed using diamond-anvil cells (DACs). The culets of the diamonds used in the DACs had diameters between 350 and 500 µm. As a gasket material, stainless-steel disks of 200 µm thickness were pre-indented to a thickness of 40 µm in which a 150–200 µm diameter hole was drilled in the center. The pressure medium was a 4:1 mixture of methanol and ethanol and the pressure was determined to an accuracy of ±0.1 GPa using the equation of state of copper [23]. The chosen pressure medium provides hydrostatic conditions to at least 10.5 GPa but affords quasi-hydrostatic conditions with increasing pressure [24]. In the four studied compounds. For each pressure point we collected two XRD patterns, one with only the sample and one with the sample and Cu. The first one was used for structural determination and the second for pressure determination. Angle-dispersive XRD experiments were performed using a monochromatic X-ray beam with a wavelength of 0.42460(8) Å focused to a spot size of 20 µm × 20 µm using multilayer Kirkpatrick–Baez mirrors. Wavelength selection was achieved through a double-crystal silicon (111) monochromator. Two-dimensional (2D) XRD patterns were collected using a Rayonix SX165 charge-coupled device which was calibrated using $LaB_6$ as a standard reference material. The detector-sample distance was 210.19 mm. The typical acquisition time was 20 seconds. The 2D diffraction images were integrated into intensity versus 2θ XRD patterns using DIOPTAS [25] and Le Bail and/or Rietveld refinements were performed with the FullProf suite [26]. For the structural analysis, we first modelled the background with a Chebyshev polynomial function of first kind with six coefficients. The shape of the peaks was modelled using a pseudo-Voigt function and Cagliotti coefficients. Given that the atomic displacement factors are more sensitive to background determination than the positional parameters, they were constrained to $B = 0.025$ Å$^2$, where $B$ is the overall displacement factor.

*2.2 Computer simulations*

Ab initio density-functional theory calculations were performed using the Vienna ab initio Simulations Package (VASP) [27] using the projector-augmented wave pseudopotentials [28] (PAW) available in the VASP database. The Ca, Zr, Ti, and O valence shell electrons configuration were [Ar] $4s^2$, [Kr] $4d^2 5s^2$, [Ar] $3d^2 4s^2$, and [He] $2s^2 2p^4$, respectively. To ensure high accuracy and well-converged results an energy cutoff of 600 eV was used. The exchange-correlation energy was described within the generalized-gradient approximation (GGA) with the Armiento and Mattsson (AM05) prescription which has an excellent performance for solids [29]. To ensure a robust convergence, a dense Monkhorst-Pack scheme was employed to discretize the Brillouin-zone (BZ) integrations [30] with a 4 x 4 x 2 k-points mesh. The self-consistency energy convergence criteria were set to $10^{-6}$ eV and the maximum force between atoms below 0.004 eV/Å, with the deviation of the diagonal tress tensor smaller than 0.1 GPa to ensure hydrostatic conditions in the simulations. The phonon calculations at the zone center of the BZ were performed using the Phonopy package [31] that provides the frequencies and eigenvectors obtaining the Raman and Infrared modes. The Phonopy package also allows to study of the dynamical stability, in this study we used a 2x2x2 supercell to obtain the phonon dispersion along the high symmetry points of the BZ.

### 3. Results and discussion

*3.1 Crystal structure of zirconolite-2M (zirconolite-2TR)*

The crystal structure of zirconolite-2M has previously been determined to have a monoclinic lattice (space group *C*2/*c*) with unit cell parameters of: *a* = 12.4413(2) Å, *b* = 7.2700(1) Å, *c* = 11.3744(2) Å, and $\beta$ = 100.557(2)° (*Z* = 8) [10], as reported under Inorganic Crystal Structure Database entry 32339. In this crystal structure, one quarter of the Ti atoms have been assigned to be at a Wyckoff 8f position with a statistical occupation of 50% [10]. Using this crystal structure as a starting model, here, the structure was refined by the Rietveld method against the XRD pattern obtained at the lowest pressure (0.1 GPa) at ambient temperature. The refinement is shown in Fig. 1(a). However, the obtained R-factors and $\chi^2$ of the fit ($R_p$ = 21.3%, $R_{wp}$ = 22.9%, $\chi^2$ = 3.23) are slightly larger than those desired for a good fit of a model to the data [32]. In addition, the refinement does not produce a smooth difference curve for the three strongest reflections (see for example the reflection around 8° in Fig. 1(a)).

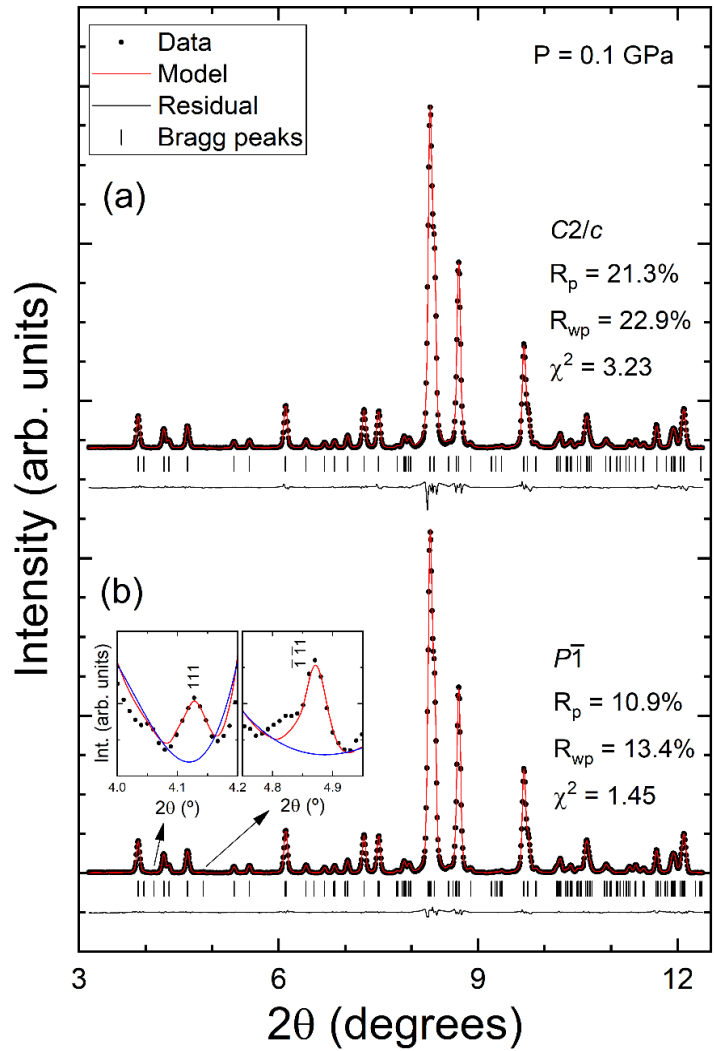

**Figure 1:** Powder X-ray patterns from zirconolite-2M at 0.1 GPa. (a) The Rietveld refinement assuming the monoclinic structure (space group *C2/c*). (b) The Rietveld refinement assuming the triclinic structure (space group $P\bar{1}$). Black circles (red lines) are the data from experiments (refinements). The vertical ticks are the calculated positions of the Bragg reflections. The black lines are the residuals. The R-values and $\chi^2$ of the refinement are given in the figure. The two insets show weak reflections at low angles that cannot be explained by the monoclinic structure. The symbols in the insets represent the experimental data, and the red (blue) lines are the modelled XRD pattern assuming the triclinic (monoclinic) structure. The Miller indices of the reflections from the triclinic structure are included.

An improved Rietveld fit can be obtained by reducing the crystal symmetry from monoclinic to triclinic, as shown in Fig. 1(b). In this case, the goodness-of-fit parameters are significantly lower than in the monoclinic case $R_p$ = 10.9%, $R_{wp}$ = 13.4%, and $\chi^2$ = 1.45. The starting model of the triclinic structure was obtained using the result from the refinement of the original monoclinic structure (space group *C2/c*, shown in Fig. 1(a)) and by using the $P\bar{1} \subset C2/c$ group-subgroup transformation. The monoclinic unit cell is related to the triclinic unit cell by applying the transformation matrix

$$\begin{pmatrix} 1/2 & -1/2 & 0 \\ 0 & 1 & 0 \\ 0 & 0 & 1 \end{pmatrix}.$$ It should be noted that the triclinic structure (space group $P\bar{1}$) not only leads to smaller residuals in the fits, but it also explains weak reflections in the XRD pattern that cannot be indexed by the monoclinic structure (space group $C2/c$). Two examples of reflections that can only be explained by the triclinic structure are shown in the insets of Fig. 1(b). Interestingly, in the triclinic structure all Ti atoms fully occupy the Wyckoff positions and there are no partially occupied positions as in the previously reported monoclinic structure. The unit cell parameters determined for the triclinic structure (space group $P\bar{1}$) at 0.1 GPa are: $a$ = 7.203(3) Å, $b$ = 7.216(3) Å, $c$ = 11.372(5) Å, $\alpha$ = 80.99(1)°, $\beta$ = 80.86(1)°, and $\gamma$ = 60.59(1)° ($Z$ = 4). The crystallographic data obtained in the course of this work for each of the zirconolite structures studied can be found in Table 1. The full description of the triclinic crystal structure can be found in the CIF file deposited in the Cambridge Crystallographic Data Centre (CCDC) under deposition number 2246629. The atomic positions can be found in Table S1 in the Supplementary Material.

| Phase | Composition | Space Group | Unit-cell parameters |
|---|---|---|---|
| zirconolite-2TR (LP, 0.1 GPa) | CaZrTi$_2$O$_7$ | $P\bar{1}$ | $a$ = 7.203(3) Å, $b$ = 7.216(3) Å, $c$ = 11.372(5) Å, $\alpha$ = 80.99(1)°, $\beta$ = 80.86(1)°, $\gamma$ = 60.59(1)° ($Z$ = 4) |
| zirconolite-2M (HP, 14.7 GPa) | CaZrTi$_2$O$_7$ | $C2/c$ | $a$ = 12.035(9) Å, $b$ = 6.987(4) Å, $c$ = 11.304(9) Å, $\beta$ = 100.12(2)° ($Z$ = 8) |
| zirconolite-4M (0.1 GPa) | Ca$_{0.75}$Zr$_{0.75}$Dy$_{0.50}$Ti$_2$O$_7$ | $C2/c$ | $a$ = 12.465(1) Å, $b$ = 7.189(1) Å, $c$ = 22.994(2) Å, $\beta$ = 84.82(1)° ($Z$ = 16) |
| zirconolite-3O (0.1 GPa) | Ca$_{0.20}$Nd$_{0.80}$ZrTi$_{1.20}$Fe$_{0.80}$O$_7$ | $Cmce$ | $a$ = 10.131(2) Å, $b$ =14.049(2) Å, $c$ = 7.322(1) Å ($Z$ = 8) |
| zirconolite-3T (0.3 GPa) | Ca$_{0.65}$Ce$_{0.35}$ZrTi$_{1.30}$Fe$_{0.70}$O$_7$ | $P3_121$ | $a$ = 7.295(2) Å, $c$ = 17.337(3) Å ($Z$ = 6) |

**Table 1:** Experimental unit cell parameters for the different zirconolite structures studied in this work as obtained from powder XRD experiments at different pressures. Atomic positions of all structures are reported in the Supplementary Material.

The triclinic zirconolite-2M structure as obtained by Rietveld refinements is shown in Fig. 2. The structure consists of planes of corner sharing CaO$_8$ and ZrO$_7$ polyhedral units running perpendicular to the [001] direction separated by HTB-type layers of Ti atoms. The Ti atoms have two different coordination polyhedra forming TiO$_6$ octahedral units and a fivefold coordinated TiO$_5$ trigonal bipyramid. We hypothesise that the observation of the monoclinic structure (space group $C2/c$) in a previous study [10] and the triclinic structure (space group $P\bar{1}$) in the present study could be related to the existence of a metastable structure. Obtaining one or the other structure could be related to differences in the specific composition and thermal history of the materials studied. Further studies are needed to clarify this issue. Since in the name Zirconolite-2M the M indicate a monoclinic symmetry, to avoid confusions for the rest of the paper we will name the triclinic structure as zirconolite-2TR. We decided to use TR for triclinic because usually T is used for tetragonal zirconolite structures.

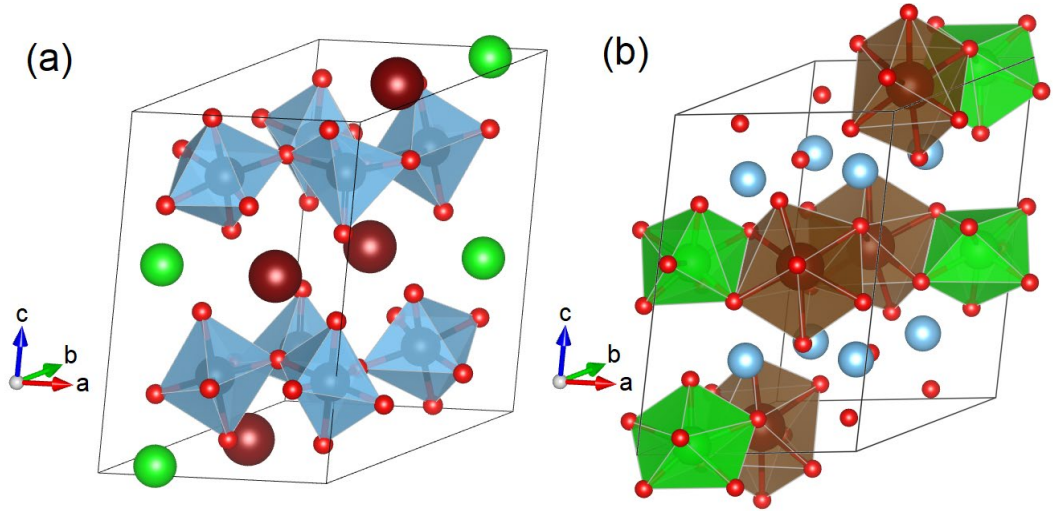

**Figure 2:** Triclinic crystal structure of zirconolite-2M. (a) The layers of $TiO_6$ and $TiO_5$ polyhedra are shown in blue and the Ca and Zr atoms are shown in brown and green respectively. (b) The $CaO_8$ and $ZrO_7$ polyhedra are shown in brown and green, respectively, and the Ti atoms are shown in blue. Oxygen atoms are represented in red.

To further investigate the stability of both phases (2M and 2TR), we have performed DFT calculations which, as will be discussed next, provide strong evidence in favour of the triclinic structure over the monoclinic one. After optimizing the two crystal structures at 0 GPa, we have calculated their phonon dispersions, which are shown in Fig. 3. We found that the monoclinic 2M structure (space group $C2/c$) has imaginary frequencies, which are plotted as negative branches in Fig. 3(a). This fact reflects the dynamical instability of the monoclinic 2M structure. In contrast, for the triclinic 2TR structure (space group $P\bar{1}$) we found that all phonon branches are positive (see Fig. 3(b)) supporting the fact that this structure is dynamically stable. Calculations of the enthalpy as a function of pressure, shown in Figure 4, indicate that the triclinic 2TR structure has a lower enthalpy than the monoclinic 2M structure for pressures below 28 GPa, making it the most thermodynamically stable structure of the two. Based on these facts, combined with the Rietveld refinements, we are confident the crystal structure of

zirconolite-2M could be redefined as triclinic. High resolution neutron diffraction experiments will be required to definitively confirm the triclinic crystal structure reported here.

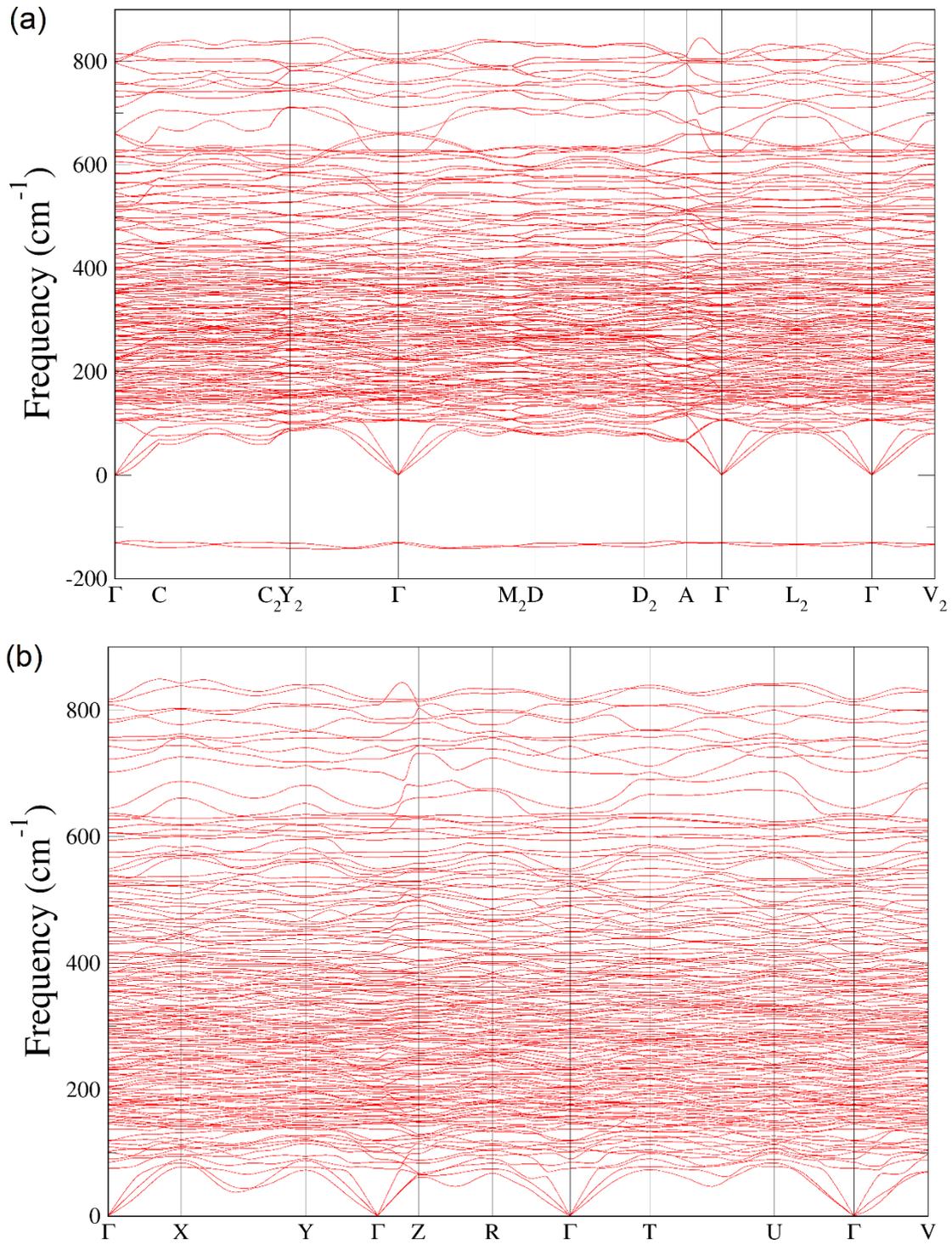

**Figure 3:** (a) Calculated phonon dispersion for the monoclinic zirconolite-2M structure (space group *C*2/*c*). (b) Calculated phonon dispersion for the proposed triclinic zirconolite-2TR structure (space group *P*$\bar{1}$).

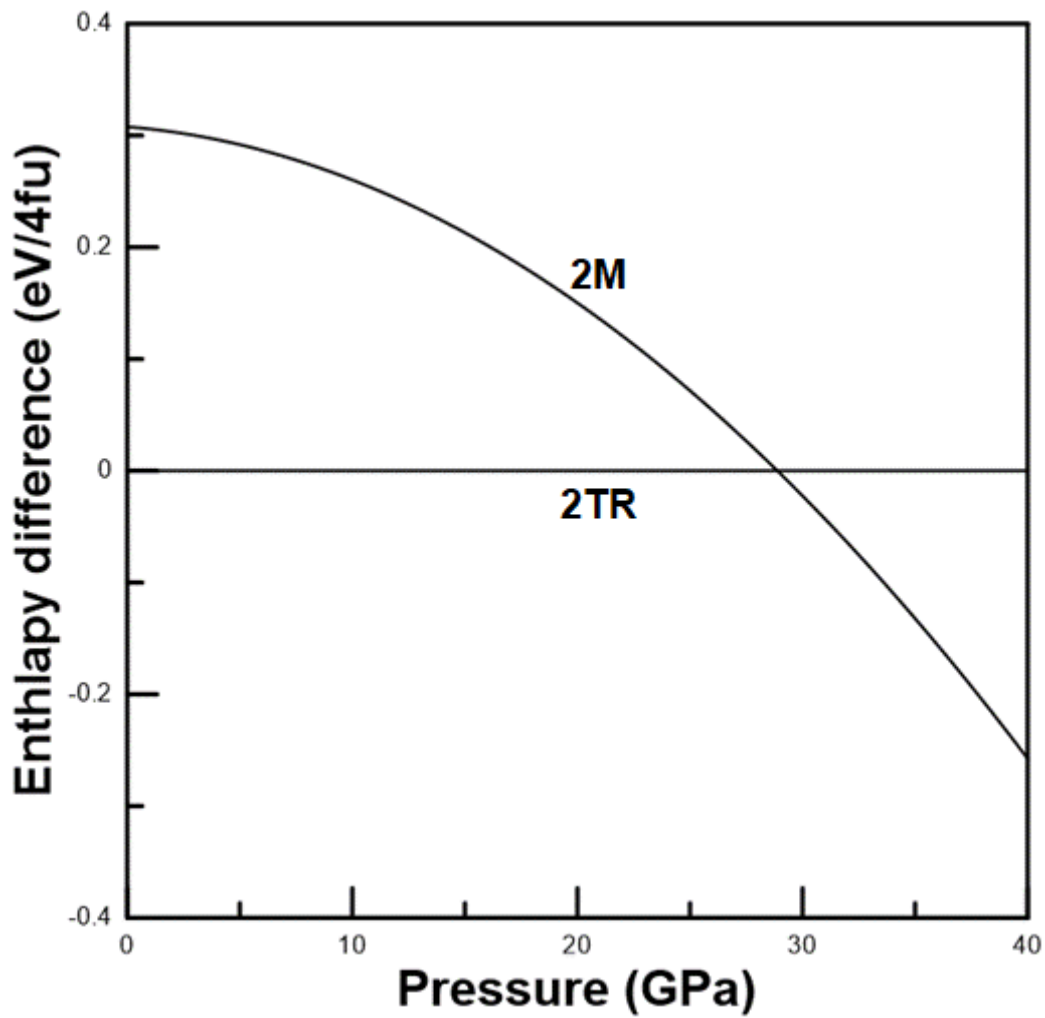

**Figure 4:** The calculated enthalpy difference per four formula units between the triclinic zirconolite-2TR and monoclinic zirconolite-2M structures using the triclinic structure as the reference (x axis).

To complete the description of the ambient-pressure triclinic structure of zirconolite-2TR we will report the calculated Raman-active and infrared (IR)-active phonon. According to group-theory, zirconolite-2TR has 66$A_g$ + 66$A_u$ phonons at the Γ-point of the Brillouin zone. Among these modes, 3 $A_u$ are acoustic modes. The optical modes are 66 $A_g$ Raman-active modes and 63 $A_u$ IR-active modes. The calculated wavenumbers of these modes are reported in Table 2. Unfortunately, there are no experiments to compare with. We decided to report the results of our calculations for comparison with future studies. Notice that the results of calculations could be quite helpful for the assignment of Raman and IR modes in a compound with a large number of phonons like zirconolite-2TR. The Raman spectrum is composed of a group of modes at low frequencies (74.5 – 119.4 cm$^{-1}$), a continuum of modes from 145.4 to 497.1 cm$^{-1}$ which in experiments should appear as a broad band, two clusters of modes in the 520.9-575.3 cm$^{-1}$ and 605.5-636.6 cm$^{-1}$ regions, and five isolated modes above 700 cm$^{-1}$. The IR spectrum has a similar frequency distribution, with a group of modes at low frequency from 84.7 to 114.3, a continuum of modes from 137 to 545.2 cm$^{-1}$, four modes from 593.7 to 644.3 cm$^{-1}$ and five modes above 722 cm$^{-1}$. The highest frequency Raman and IR modes originate from Ti-O stretching vibrations. Their frequencies are comparable to those of the Ti-O stretching modes of the TiO$_6$ octahedron in calcium titanate and lead zirconium titanate [33, 34], two compounds with Ti coordination polyhedra similar to those in zirconolite-2M. The lowest frequency modes can be associated to vibrations between Ca or Zr and the TiO$_6$ octahedra. The rest of modes involve more complex vibrations, with some of them being Ti-O-Ti and O-Ti-O bending vibrations.

| Mode | ω(cm$^{-1}$) | Mode | ω(cm$^{-1}$) | Mode | ω(cm$^{-1}$) | Mode | ω(cm$^{-1}$) |
|---|---|---|---|---|---|---|---|
| A$_g$ | 74.5 | A$_g$ | 334.9 | A$_u$ | 84.7 | A$_u$ | 347.8 |
| A$_g$ | 89.6 | A$_g$ | 338.9 | A$_u$ | 100.5 | A$_u$ | 353.1 |
| A$_g$ | 102.4 | A$_g$ | 355.2 | A$_u$ | 108.8 | A$_u$ | 367.4 |
| A$_g$ | 118.3 | A$_g$ | 363.5 | A$_u$ | 114.3 | A$_u$ | 358.1 |
| A$_g$ | 119.4 | A$_g$ | 364.0 | A$_u$ | 136.6 | A$_u$ | 372.0 |
| A$_g$ | 145.4 | A$_g$ | 370.8 | A$_u$ | 147.4 | A$_u$ | 383.2 |
| A$_g$ | 147.7 | A$_g$ | 378.8 | A$_u$ | 152.1 | A$_u$ | 386.3 |
| A$_g$ | 156.5 | A$_g$ | 386.7 | A$_u$ | 165.6 | A$_u$ | 392.1 |
| A$_g$ | 158.4 | A$_g$ | 399.5 | A$_u$ | 166.9 | A$_u$ | 414.1 |
| A$_g$ | 168.7 | A$_g$ | 405.7 | A$_u$ | 175.4 | A$_u$ | 429.9 |
| A$_g$ | 174.3 | A$_g$ | 411.2 | A$_u$ | 177.2 | A$_u$ | 440.1 |
| A$_g$ | 175.0 | A$_g$ | 416.8 | A$_u$ | 183.5 | A$_u$ | 449.6 |
| A$_g$ | 183.4 | A$_g$ | 434.9 | A$_u$ | 187.3 | A$_u$ | 465.3 |
| A$_g$ | 193.3 | A$_g$ | 448.7 | A$_u$ | 195.1 | A$_u$ | 470.9 |
| A$_g$ | 200.0 | A$_g$ | 455.7 | A$_u$ | 203.3 | A$_u$ | 490.6 |
| A$_g$ | 202.5 | A$_g$ | 476.3 | A$_u$ | 215.3 | A$_u$ | 502.3 |
| A$_g$ | 205.0 | A$_g$ | 489.8 | A$_u$ | 222.7 | A$_u$ | 512.7 |
| A$_g$ | 212.9 | A$_g$ | 497.1 | A$_u$ | 230.6 | A$_u$ | 525.3 |
| A$_g$ | 225.0 | A$_g$ | 520.9 | A$_u$ | 234.0 | A$_u$ | 530.0 |
| A$_g$ | 232.2 | A$_g$ | 525.4 | A$_u$ | 244.1 | A$_u$ | 536.5 |

**Table 2:** Calculated Raman frequencies (A$_g$ modes) and IR frequencies (A$_u$ modes) of triclinic Zirconolite-2M.

*3.2 High-pressure study of zirconolite-2TR*

Fig. 5 shows a selection of powder XRD patterns measured at different pressures during compression up to 30.5 GPa and decompression to ambient pressure. As the pressure increases, all XRD patterns below 14.7 GPa can be assigned to the low-pressure triclinic phase discussed in Section 3.1. The structural information of the low-pressure phase at 13.5 GPa is reported in Table S2 of the Supplementary Material. At 14.7 GPa there are clear changes in the XRD pattern in the angular region of the two strongest reflections (8-9°). These changes indicate the occurrence of a structural phase transition.

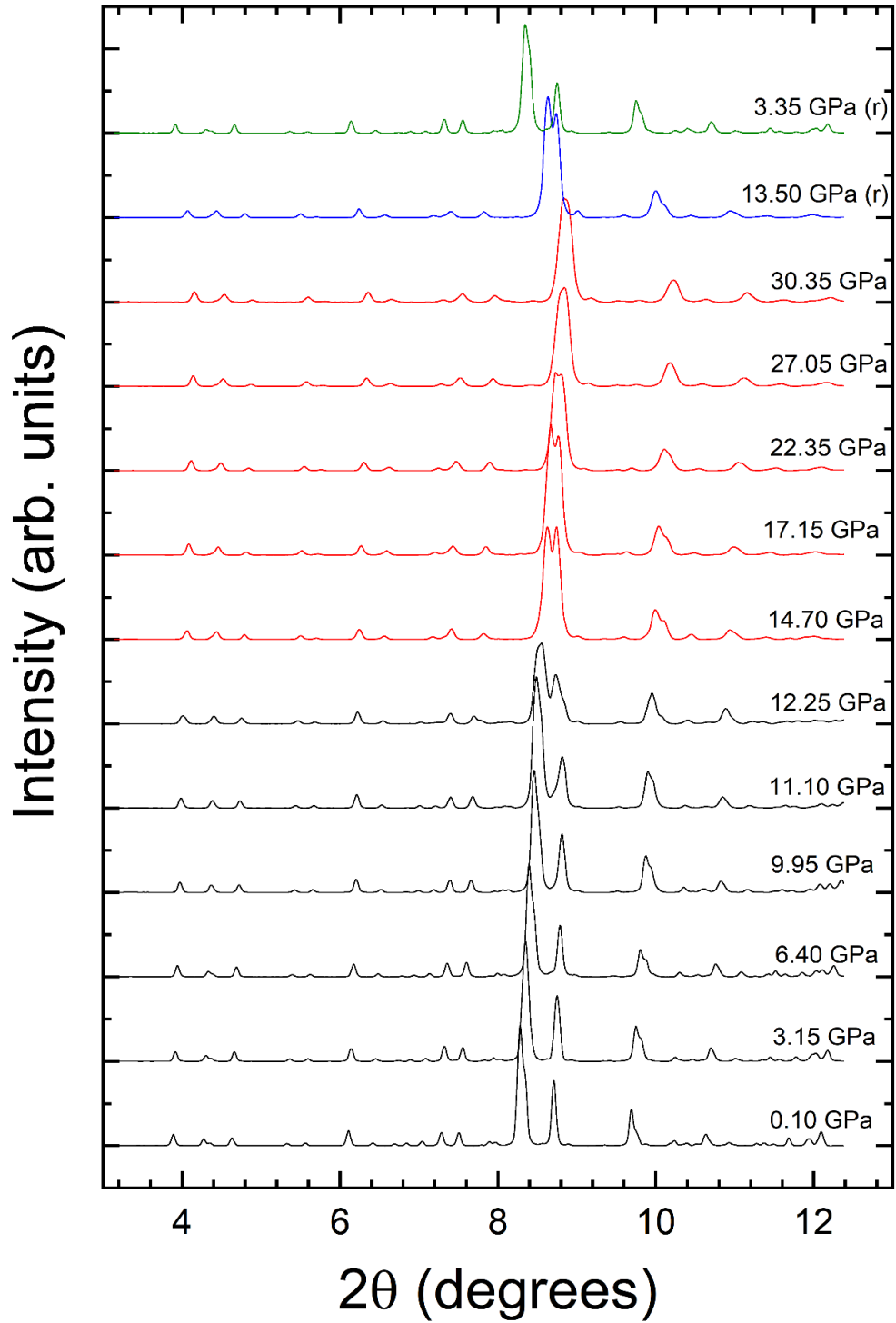

**Figure 5:** Selection of XRD patterns measured at different pressures (indicated in the figure) in zirconolite-TR. Black(red) colour is used for the low(high)-pressure phase. Blue(green)-colour is used for the high(low)-pressure phase under pressure release, indicated by (r) next to the pressure.

The transition pressure, 14.7 GPa, is just above the quasi-hydrostatic limit of the used pressure medium used [24]. It is therefore conceivable that the observed phase transition is triggered by non-hydrostatic stresses [35]. However, since no phase transition was found in any of the other three zirconolites studied here, where the same pressure medium was used, we consider that the observed transition is indeed caused by pressure and not by non-hydrostatic stresses. Furthermore, our results are in agreement with the previous study of Salamat *et al.* [20] where a phase transition was reported in zirconolite-2M at 15.6 GPa. In the experiments reported in Ref. [20] the pressure medium was helium, which remains quasi-hydrostatic up to 40 GPa [36]. The agreement between the transition pressures of the two experiments supports our hypothesis that the transition is not driven by non-hydrostatic stresses. After the phase transition, above 14.7 GPa, the most noticeable change in the XRD pattern is the merging of the two strongest reflections (see Fig. 5). However, no evidence of a second phase transition was found up to the maximum pressure studied of 30.4 GPa. On decompression, the HP phase is retained to at least 13.5 GPa. The low-pressure phase is recovered by 3.4 GPa. Unfortunately, we did not collect any XRD data on decompression between 13.5 and 3.4 GPa, so the hysteresis of the transition cannot be fully characterized given the present data.

A candidate crystal structure for the HP phase of Zirconolite 2TR is proposed here. Using DICVOL [37] we found that the indexing of the XRD pattern measured at 14.7 GPa gave the lowest figure of merit for the space group *C*2/*c*. A subsequent Le Bail fit using the space group *C*2/*c* gave the starting parameters for a Rietveld refinement, for which we used the atomic positions obtained from DFT calculations of the triclinic structure transformed to *C*2/*c* symmetry. The results of the refinement are shown in Fig. 6. The refinement converged with smooth profile residuals and good goodness-of-fit parameters of: $R_p$ = 13.3%, $R_{wp}$ = 19.4%, and $\chi^2$ = 2.56.

The crystal structure of the HP phase is shown in Fig. 7. The structure is a symmetrized version of the low-pressure phase, which can be described as layers of edge-sharing $CaO_8$ and $ZrO_7$ polyhedra separated by layers of interconnected $TiO_6$ and $TiO_5$ polyhedra. Interestingly, the HP structure of zirconolite-2TR is isomorphous to zirconolite-2M. The unit cell parameters obtained from Rietveld refinement of the XRD pattern measured at 14.7 GPa are: *a* = 12.035(9) Å, *b* = 6.987(4) Å, *c* = 11.304(9) Å, and *β* = 100.12(2)°. The atomic positions and full crystallographic information of the HP crystal structure can be found in the CCDC under deposition number 2268224. The atomic positions of the high-pressure phase can be found in Table S3 of the Supplementary Material.

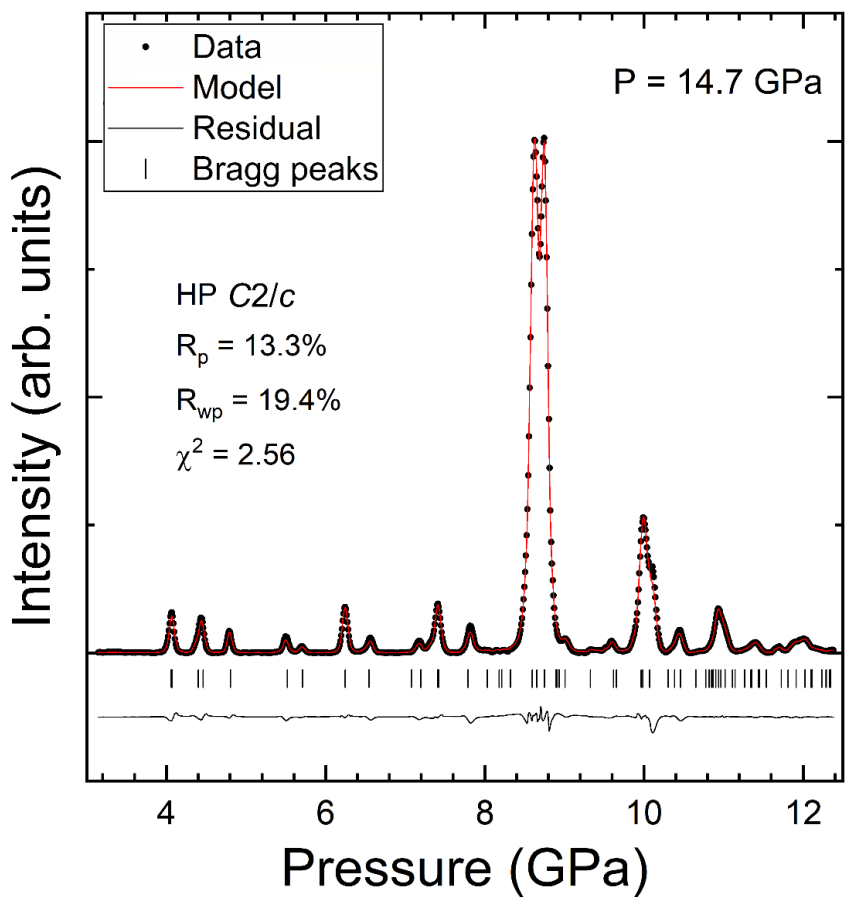

**Figure 6:** Rietveld refinement of the integrated powder X-ray diffraction pattern of the HP phase of Zirconolite-2TR at 14.7 GPa using the proposed monoclinic structure. Black circles (red lines) are the experimental data (refinement). The vertical ticks are the fitted positions of the Bragg reflections. The black lines are the residuals. The $R$-values and $\chi^2$ of the refinement are given in the figure.

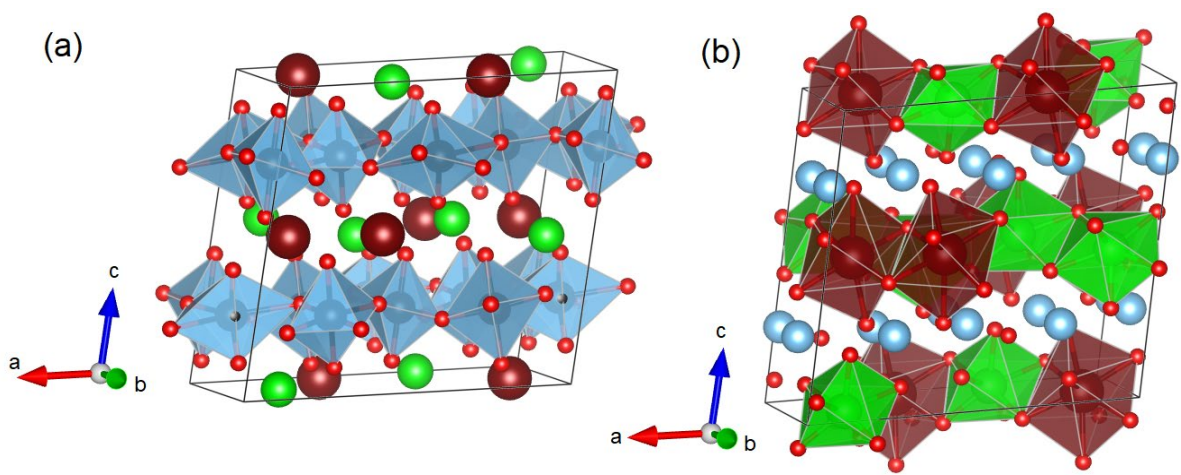

**Figure 7:** High-pressure monoclinic crystal structure of zirconolite-2TR. (a) The layers of $TiO_6$ and $TiO_5$ polyhedra are shown in blue and Ca and Zr atoms in brown and green respectively. (b) The $CaO_8$ and $ZrO_7$ polyhedra are shown in brown and green, respectively, and the Ti atoms are shown in black. Oxygen atoms are represented in red.

The proposed monoclinic crystal structure explains all the XRD patterns measured from 14.7 to 30.4 GPa. Note that the space group of the HP monoclinic structure, space group *C*2/*c*, is the same as that previously reported for the ambient pressure zirconolite-2M structure [10]. However, in the crystal structure we propose for the HP phase, the Wyckoff position for one Ti atom is different from that of its equivalent counterpart in the ambient-pressure structure [10]. In our HP structure we have one Ti atom at 8f and two Ti atoms at 4e where all sites have an occupation of 1 (see file number 2268224 in the CCDC), whereas in the previously reported ambient pressure structure [10] two Ti atoms are at 8f (one with occupation 1 and one with occupation 1/2) and one Ti atom is at 4e. In summary, the two monoclinic structures are similar but they are not isomorphic. The unit cell parameters of our HP structure are similar to those of the previously reported ambient pressure phase, but with smaller values for *a*, *b*, and *c* due to the pressure induced contraction of the unit-cell at 14.7 GPa. The assignment of the crystal structure of the HP phase to the monoclinic structure described by the space group *C*2/*c* agrees with the conclusions of our DFT calculations, which support that the monoclinic structure has a lower enthalpy than the triclinic structure under compression (see Figure 4). In addition, the DFT calculations also show that the monoclinic HP structure reported here becomes dynamically stable after the phase transition as shown by the phonon dispersion plotted in Figure 8.

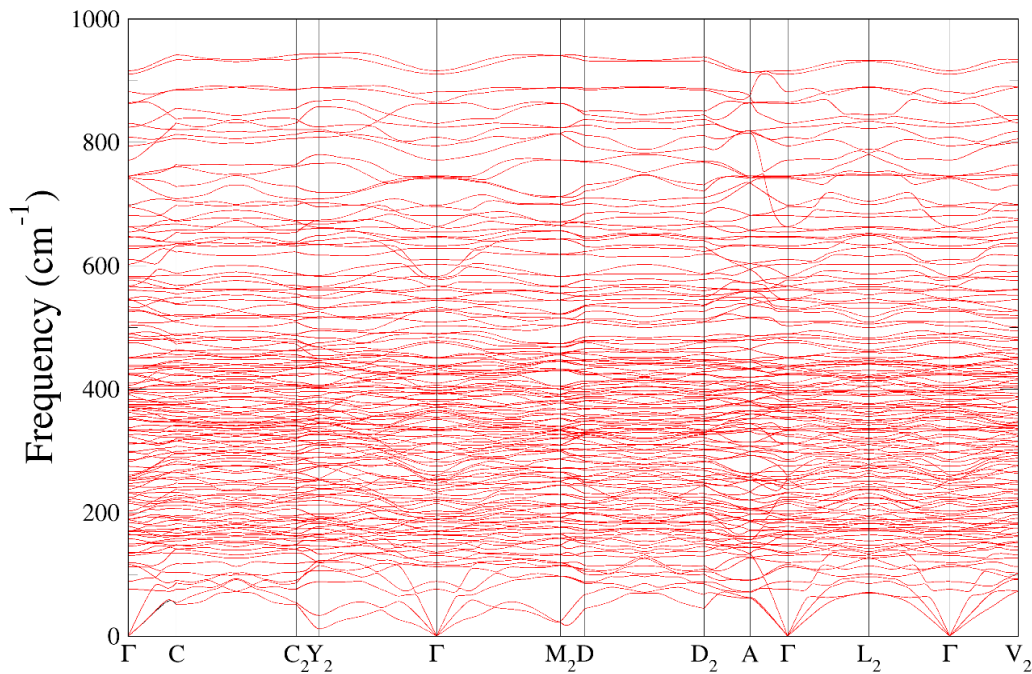

**Figure 8:** Phonon dispersion calculated at 23 GPa for the high-pressure monoclinic structure of zirconolite-2TR (space group *C*2/*c*) showing that the structure is dynamically stable.

A diagram summarizing the results of the present study and the previous study [20] is shown in Fig. 9, where it can be seen that although both studies found a similar transition pressure (14.7 and 15.6 GPa), there are some discrepancies regarding the crystal structure of the HP phase. The monoclinic HP crystal structure proposed here has a different unit-cell and space group than that proposed by Salamat *et al.* [20], who did not fully solve the crystal structure. Based on a Le Bail fit, they have proposed a different monoclinic structure that could be described by space groups $P2_1$ or $P2_1/m$. However, their attempt to solve the structure was carried out with data collected at 40 GPa, which is beyond the pressure limit covered by our study. Therefore, we hypothesise that the finding of different structures in different pressure ranges ($C2/c$ from 14.7 to 30.5 GPa and $P2_1$ or $P2_1/m$ from 40 to 56 GPa) could indicate the existence of a second phase transition at pressures higher than those covered by the present study, with the second HP phase ($P2_1$ or $P2_1/m$) being stable at 40 GPa.

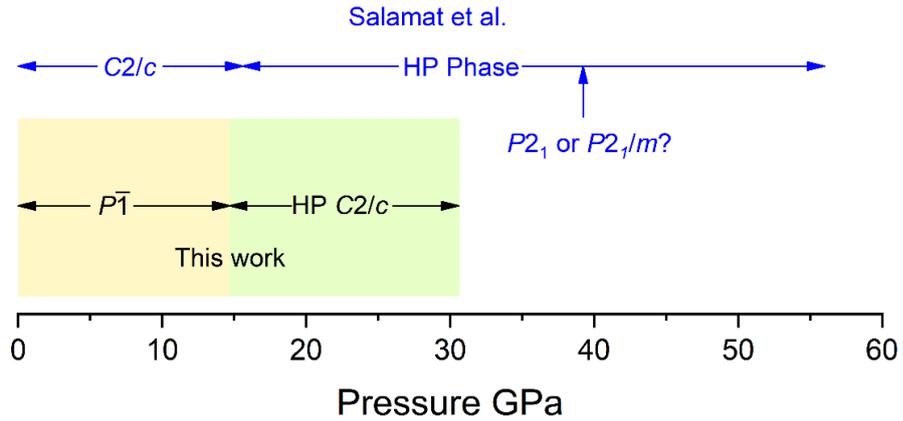

**Figure 9:** Schematic representation of the HP structural sequence of zirconolite-2M according to this work and the previous HP study [20].

| Phase | Composition | $V_0$ (Å$^3$) | $K_0$ (GPa) | $K'_0$ |
|---|---|---|---|---|
| zirconolite-2TR LP (exp.) | CaZrTi$_2$O$_7$ | 506.4(8) | 170(3) | 4.1(6) |
| zirconolite-2TR LP (DFT) |  | 505.01(9) | 166.1(7) | 4.9(1) |
| zirconolite-2M HP (exp.) | CaZrTi$_2$O$_7$ | 1006(5) | 171(3) | 4.0(9) |
| zirconolite-2M HP (DFT) |  | 1005.6(2) | 160(1) | 4.3(1) |
| zirconolite-4M (exp.) | Ca$_{0.75}$Zr$_{0.75}$Dy$_{0.50}$Ti$_2$O$_7$ | 2053(1) | 164(3) | 3.7(1) |
| zirconolite-3O (exp.) | Ca$_{0.20}$Nd$_{0.80}$ZrTi$_{1.20}$Fe$_{0.80}$O$_7$ | 1040(1) | 163(10) | 11(2) |
| zirconolite-3T (exp.) | Ca$_{0.65}$Ce$_{0.35}$ZrTi$_{1.30}$Fe$_{0.70}$O$_7$ | 790.0(3) | 159.3(3) | 8.1(3) |

**Table 3:** Zero-pressure volume, bulk modulus and its zero-pressure derivative for the zirconolite phases studied in this work as obtained from fits to the experimental (exp.) or theoretical (DFT) P-V curves with a third-order Birch-Murnaghan (BM-3) equation of state (EOS).

From the XRD patterns we obtained the pressure dependence of the unit cell parameters of the low- and high-pressure phases of zirconolite-2TR. The results are shown in Fig. 10. The agreement between our DFT calculations and experiments is excellent for the low-pressure phase. The agreement is also good in the case of the volume if we compare with the results reported by Salamat *et al*. [20] for zirconolite-2M. This is not surprising because both structures have a very similar polyhedral framework. Note that only the volumes can be compared between zirconolite-2TR and zirconolite-2M because their different crystal structures.

The pressure dependence of the volume is fitted with a third-order Birch-Murnaghan equation of state (EOS) [38]. From the experiments we obtained the unit-cell volume at zero pressure $V_0$ = 506.4(8) Å$^3$, the bulk modulus at zero pressure $K_0$ = 170(3) GPa, and its pressure derivative $K_0$' = 4.1(6). From DFT calculations we obtained $V_0$ = 505.01(9) Å$^3$, $K_0$ = 166.1(7) GPa, and $K_0$' = 4.9(1). The EOS parameters for this phase and the rest of the structures studied in this work are given in Table 3. The values of the bulk modulus and its pressure derivative fall within the one sigma confidence ellipse of the experimental results. In the study by Salamat *et al*. [20] a slightly larger (smaller) $K_0$ = 188(15) GPa ($K_0'$ = 3.6(1)) was found. This is not surprising as both parameters are correlated. The increase in $K_0$ is compensated for by the decrease in $K_0'$. However, these values still fall within the one-sigma confidence ellipse of present results.

In zirconolite-2TR there is a volume discontinuity of 1% at the phase transition. Both experiments and calculations give a similar pressure dependence of the volume for the HP phase. The obtained EOS parameters for the HP phase are $V_0$ = 1006(5) Å$^3$, $K_0$ = 171(3) GPa, $K_0$' = 4.0(9) according to experiments and $V_0$ = 1005.6(2) Å$^3$, $K_0$ = 160(1) GPa, $K_0$' = 4.3(1) according to calculations. This implies that there is no change in the bulk modulus at the phase transition. This is consistent with the fact that the transition does not involve changes in the packing efficiency of the crystal structure or the formation of new bonds. When comparing the unit cell parameters between experiment and DFT calculations, the agreement is not as good as for the low-pressure phase. However, the differences are limited to less than 1.5%, except for the *b*-axis where the differences are larger. On the other hand, the pressure dependences obtained from calculations and experiments are similar. In both phases the compressibility is slightly anisotropic (see Fig. 10). For the low-pressure phase, for which the data are of higher quality because the sample conditions were quasi-hydrostatic, we determined the compressibility tensor and obtained, using the PASCaL calculator [39], the principal axes of compressibility as well as the corresponding axial compressibility. We obtained $\kappa_1$= 1.89(6) $10^{-3}$ GPa$^{-1}$, $\kappa_2$= 1.73(1) $10^{-3}$ GPa$^{-1}$, and $\kappa_3$= 1.17(4) $10^{-3}$ GPa$^{-1}$, with the corresponding principal axes of compressibility oriented along directions [$\bar{8}$71], [580], and [1$\bar{4}$9], respectively.

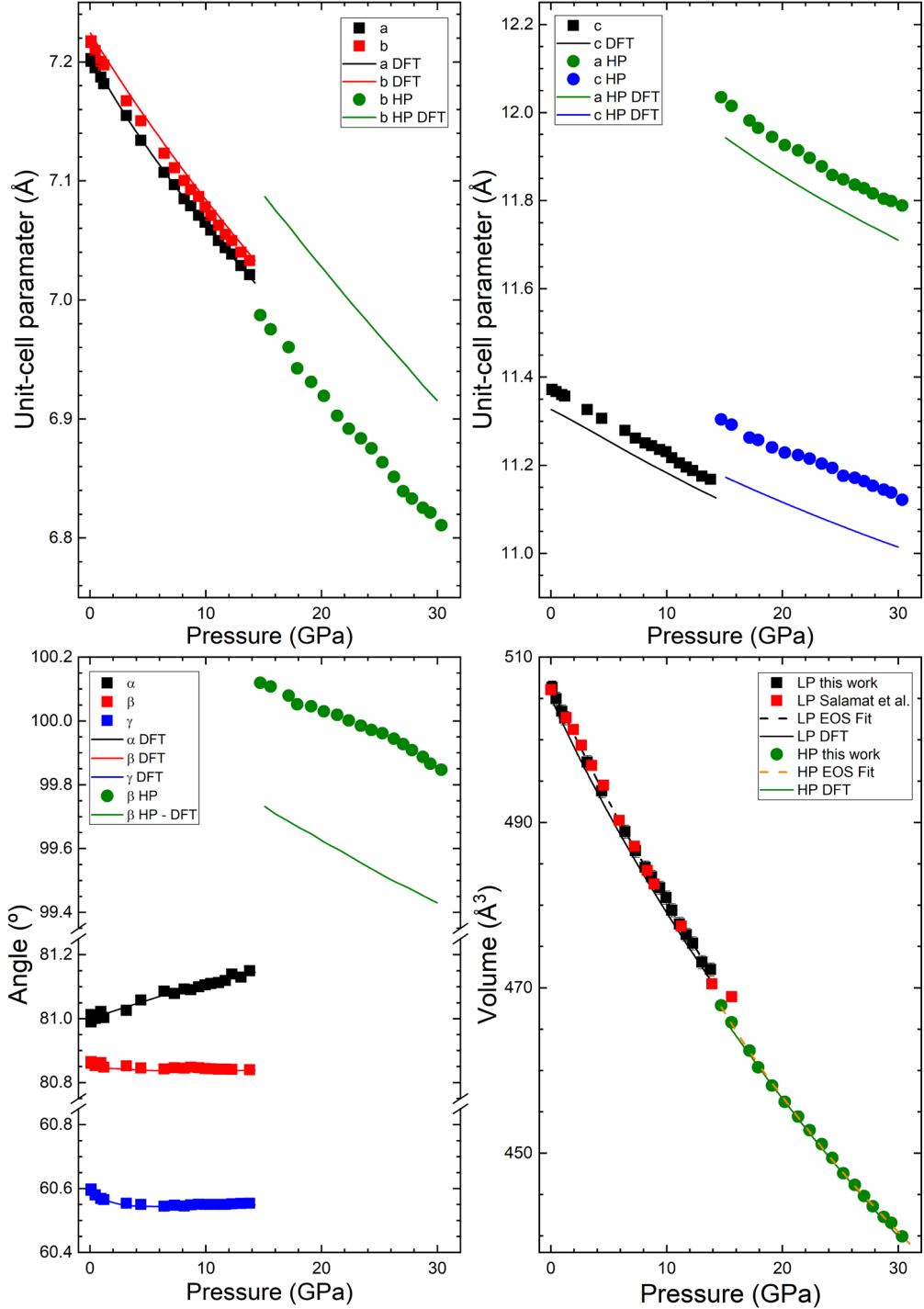

**Figure 10:** Pressure dependence of the unit cell parameters and volume of the low-pressure and high-pressure phases of zirconolite-2M. Results from experiments (symbols) and DFT calculations (solid lines) are described in the legends of the figure. For the volume of the HP phase, *V*/2 is plotted to facilitate the comparison with the low-pressure phase. The equations of state (described in the text) obtained from fits to experimental data, including those of Ref. [20], are shown as dashed lines.

*3.3 High-pressure study of zirconolite-4M*

Rietveld profiles fitted to HP XRD patterns from zirconlite-4M are shown in Fig. 11. In contrast to zirconlite-2TR, no phase transition is observed over the whole pressure range investigated in this work and all XRD patterns can be assigned to the known monoclinic structure [40] up to 29.9 GPa. The unit cell parameters obtained from the XRD pattern measured at 0.1 GPa are given in Table 1. At 29.9 GPa the unit-cell parameters are: *a* = 11.807(2) Å, *b* = 6.826(1) Å, *c* = 22.266(5) Å, and $\beta$ = 84.89(1)°. The atomic positions and the complete crystallographic information of the zirconlite-4M crystal structure can be found in Table S4 of the Supplementary Material.

The pressure dependence of unit cell parameters and unit cell volume are shown in Fig. 12. In the unit-cell parameters there is a slope change around 15 GPa, which could be related to non-hydrostatic effects [41]. The volume as a function of pressure was fitted with a third-order Birch-Murnaghan equation of state [38], yielding converged values of unit-cell volume at zero pressure of: $V_0$ = 2053(1) Å$^3$, the bulk modulus at zero pressure $K_0$ = 164(3) GPa, and its pressure derivative $K_0'$ = 3.7(1), making the compressibility of zirconlite-4M very similar to that of the LP and HP phases of zirconlite-2TR ($K_0$ = 170(3) and 171(3) GPa, respectively; see Table 3). The zirconolite-4M structure is an intergrowth of pyrochlore and zirconolite-2M [4]. The fact that zirconolites-2TR, 2M, and 4M have similar bulk moduli suggests that the polyhedral stiffness dominates the compressibility. This is consistent with the idea proposed by Hazen and Finger [42] that in certain ternary oxides the bulk modulus can be directly correlated to the compressibility of the coordination polyhedra of the larger cations. In the current case these cations are Zr and Ca (with ionic radii of 0.84 Å and 1.12 Å, respectively). The ionic radius of Ti is 0.605 Å, which makes the Ti coordination polyhedra very rigid and highly incompressible [43]. Under this hypothesis Errandonea and Manjon [44] established an empirical relationship correlating the bulk modulus with bond distances and the cation valence. Using the relationship proposed in Ref. [44] the bulk moduli of $CaO_8$ and $ZrO_7$ can be estimated to be 83 GPa and 233 GPa, respectively. Their average is 163 GPa, which is consistent with the bulk modulus we obtained for zirconolites-2TR and 4M, supporting the idea that polyhedral stiffness is the controlling factor in determining compressibility.

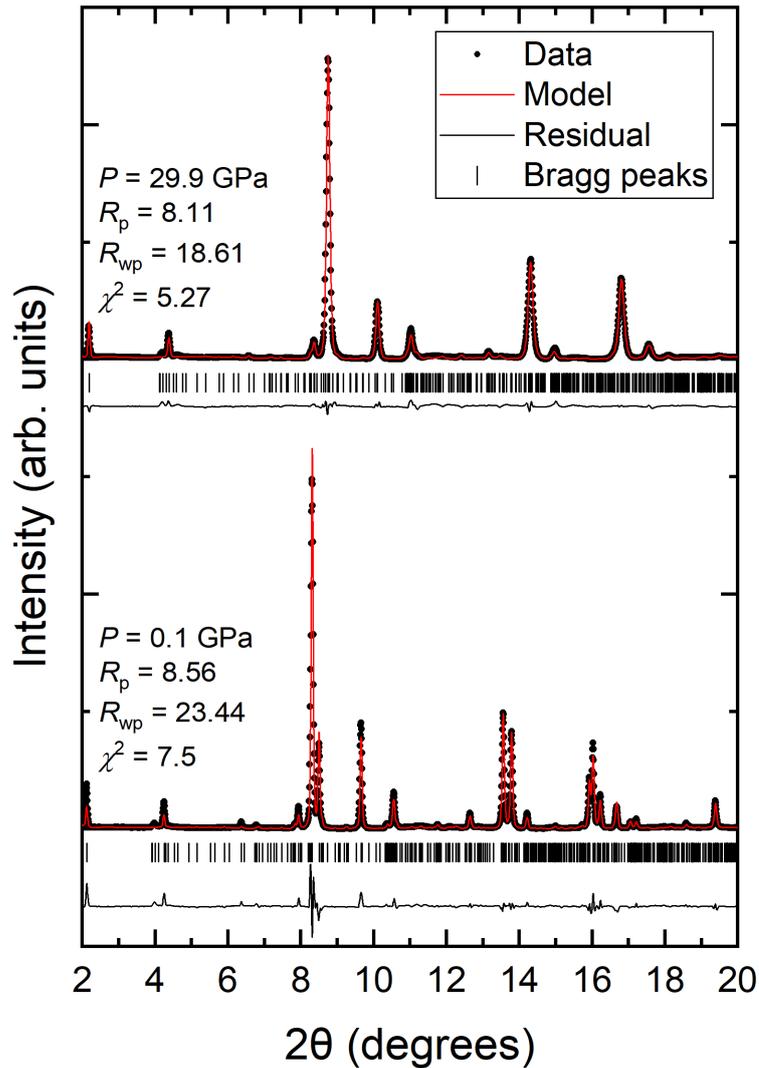

**Figure 11:** Rietveld profiles fitted to powder X-ray patterns from zirconolite-4M at 0.1 GPa and 29.9 GPa. Black circles (red lines) are the experimental data (refinements). Vertical ticks are the calculated positions of Bragg reflections. The black lines are the residuals. The R-values and $\chi^2$ of the refinements are given in the figure.

The compressibility tensor was obtained using the PASCaL web tool for Principal Axis Strain Calculations [39], yielding the principal compression axes with magnitudes: $\kappa_1 = 1.76(4)\ 10^{-3}$ GPa$^{-1}$, $\kappa_2 = 1.69(2)\ 10^{-3}$ GPa$^{-1}$, and $\kappa_3 = 0.98(2)\ 10^{-3}$ GPa$^{-1}$. The directions of the principal axes of compressibility are [100], [010], and [107], respectively. The major and intermediate compression axes, $\kappa_1$ and $\kappa_2$, are exactly parallel to the crystallographic $a$ and $b$-axes respectively. The major and intermediate compression axes are almost twice as compressible as the minor compression axis, $\kappa_3$, indicating a significant anisotropic compressibility in zirconlite-4M. This can be seen approximately in Fig. 12 whereby the change in the crystallographic $a$ axis (~0.7 Å) is approximately twice that of the change in the $b$ axis (~0.4 Å) over the pressure range studied. This is consistent with an intergrowth structure such as that of Zirconlite-4M.

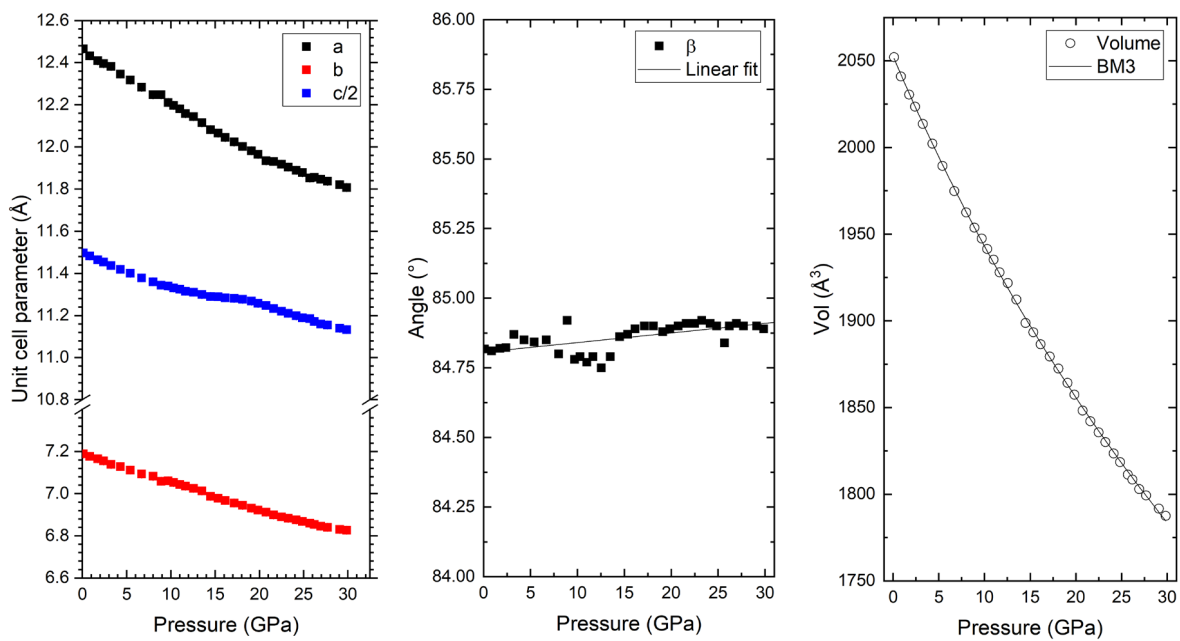

**Figure 12:** Pressure dependence of the unit-cell parameters and volume of zirconolite-4M determined from Rietveld refinement of experimental XRD data. The lattice parameter *c* is divided by a factor of 2 in the figure to facilitate comparison. EOS are shown as indicated in the legends in the figure. BM3 refers to third order Birch-Murnaghan EOS used to fit the volume.

*3.4 High-pressure study of zirconolite-3O*

Figure. 13 shows a selection of HP XRD patterns from zirconlite-3O up to 22.2 GPa. In this case, all patterns can be assigned to the known orthorhombic structure. Complete structural information of the structure can be found in Table S5 of the Supplementary Material. The patterns from this sample showed a preferred orientation, therefore Le Bail analyses were performed to extract unit-cell parameters, rather than Rietveld refinements as in the case of the previous two samples. There is no evidence of phase transition up to the highest pressure covered by the present experiments. However, there is a progressive broadening with pressure of the reflections with pressure. In Figure 14 we present the pressure dependence of unit cell parameters and volume as obtained with the Le Bail fits. The unit cell parameters obtained from the XRD patterns measured at 0.1 GPa are: $a$ = 10.131(2) Å, $b$ =14.049(2) Å, and $c$ = 7.322(1) Å. In this case, the volume as a function of pressure was fitted with a third-order Birch-Murnaghan equation of state, which allowed us to obtain converged values for the unit cell volume at zero pressure $V_0$ = 1040(1) Å$^3$, and a zero-pressure bulk modulus of $K_0$ = 163(10) GPa with a pressure derivative of $K_0'$ = 11(2). The resulting bulk modulus is comparable to that obtained for the zirconolite phases discussed above (see Table 3).

However, the $K_0'$ value is considerably larger, suggesting that this phase is less compressible than the other structures. If a second-order Birch-Murnaghan equation of state is considered in this case, the resulting zero-pressure bulk-modulus is as large as $K_0$ = 218(4) GPa (with a fixed $K_0'$ = 4). However, it is likely that the observed reflection broadening, which is particularly important above ~15 GPa and could be caused by non-hydrostatic effects, affects the quality of the EOS fits [45]. We also highlight a change in the slope of unit-cell parameter near 15 GPa (see Fig. 14), that is also indicative of the influence of non-hydrostatic stresses. Using the PASCaL tool [39] to extract the principal compression axes of the 3O structure from the pressure dependence of the lattice parameters, we found $\kappa_1$= 1.42(3)·10$^{-3}$ GPa$^{-1}$, $\kappa_2$= 0.90(3)·10$^{-3}$ GPa$^{-1}$, and $\kappa_3$= 1.01(3)·10$^{-3}$ GPa$^{-1}$ in the orthorhombic directions [001], [010] and [100], respectively. In this case, two axes show a comparable compressibility while the third axis, parallel to the *c*-axis, is about 50% more compressible.

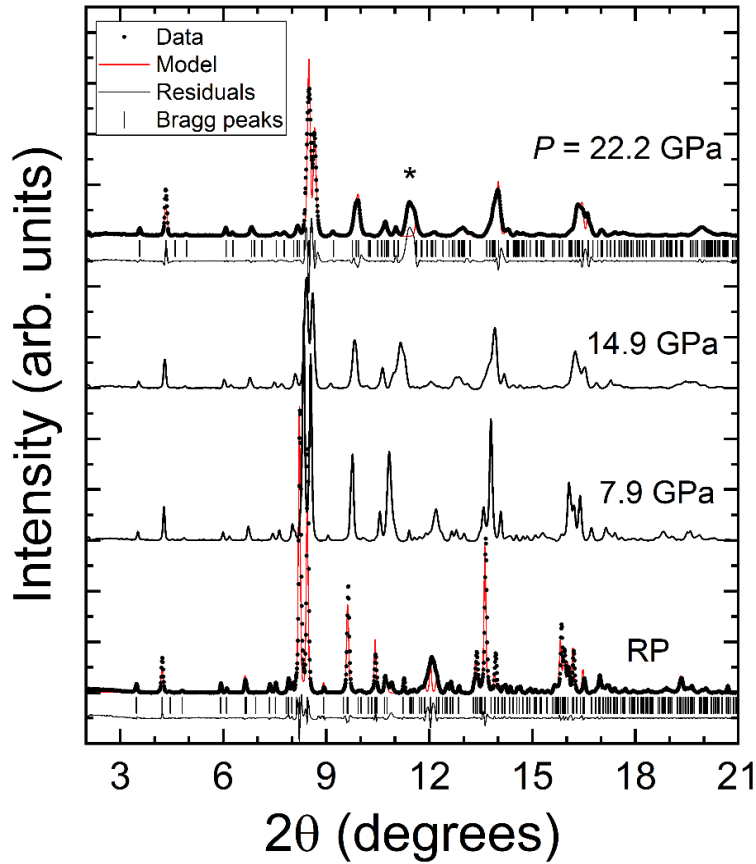

**Figure 13:** Selection of XRD patterns measured at different pressures (indicated in the figure) for zirconolite-3O. The top and bottom scans include Le Bail fits (in red) together with their residuals and the positions of the Bragg reflections. The asterisk indicates scattering that can be attributed to the steel gasket used in the experiments, which appeared in some of the diffraction patterns.

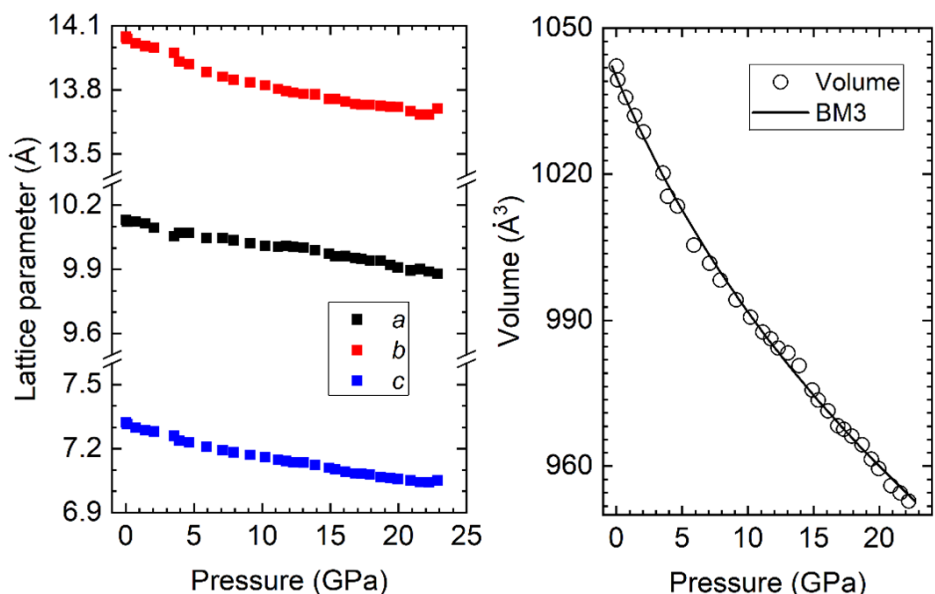

**Figure 14:** Pressure dependence of the unit-cell parameters and volume of zirconolite-3O determined from Le Bail fits to the experimental XRD data. A third-order Birch-Murnaghan EOS was used to fit the volume data.

*3.5 High-pressure study of zirconolite-3T*

Figure 15 shows a selection of XRD patterns measured in zirconolite-3T up to 29.7 GPa. A complete description of the crystal structure of zirconolite-3T can be found in Table S6 of the Supplementary Material. As with zirconolite-3O, these data cannot be refined using the Rietveld method due to the preferential orientation of the grains. Instead, a Le Bail analysis was performed to extract the unit-cell parameters. The values thus obtained at the lowest measured pressure (0.3 GPa) are the following: $a$ = 7.295(2) Å, and $c$ = 17.337(3) Å. The pressure dependence of the unit-cell parameters and volume are provided in Figure 16. A third-order BM EOS [38] is fitted to the volume data (Fig. 16), giving a zero-pressure unit cell volume $V_0$ = 790.0(3) Å$^3$, and a zero-pressure bulk-modulus of $K_0$ = 159.3(3) GPa with a pressure derivative of $K_0'$ = 8.1(3). As can be seen in Table 3, the compressibility behaviour found for this structure is also consistent with that found for the rest of the zirconolite phases.

From the lattice parameters extracted from the zirconolite-3T structure, the principal compression axes were obtained using the PASCaL calculator [39]. The resulting values are $\kappa_1 = \kappa_2$ = 1.19(4)·10$^{-3}$ GPa and $\kappa_3$ = 0.98(1)·10$^{-3}$ GPa in the directions [100]/[010] and [001], respectively. It is found that the compression axis $\kappa_3$ coincides with the crystallographic $c$-axis. For this crystal structure, all components of the compression axis have comparable values, with only a ~20% reduction in the case of $\kappa_3$. This observation indicates that, unlike the rest of Zirconolite structures, the 3T structure exhibits a fairly isotropic compressibility, probably due to the higher symmetry of its crystal structure.

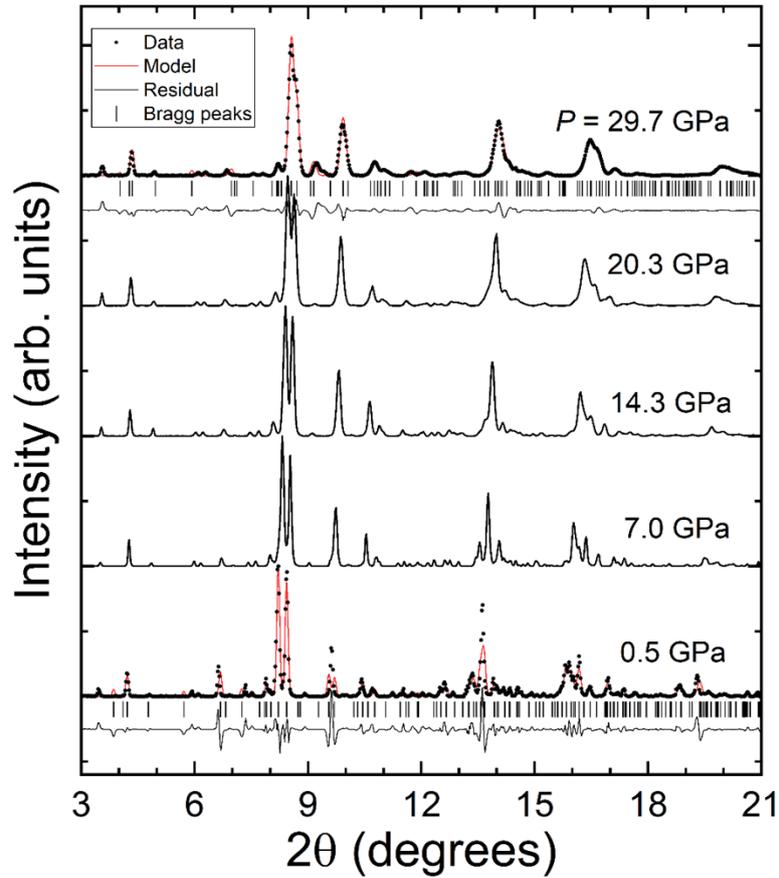

**Figure 15:** Selection of XRD patterns measured at different pressures (indicated in the figure) for zirconolite-3T. Highest and lowest pressure patterns include Le Bail fitted curves (in red) together with their residuals and Bragg reflection positions.

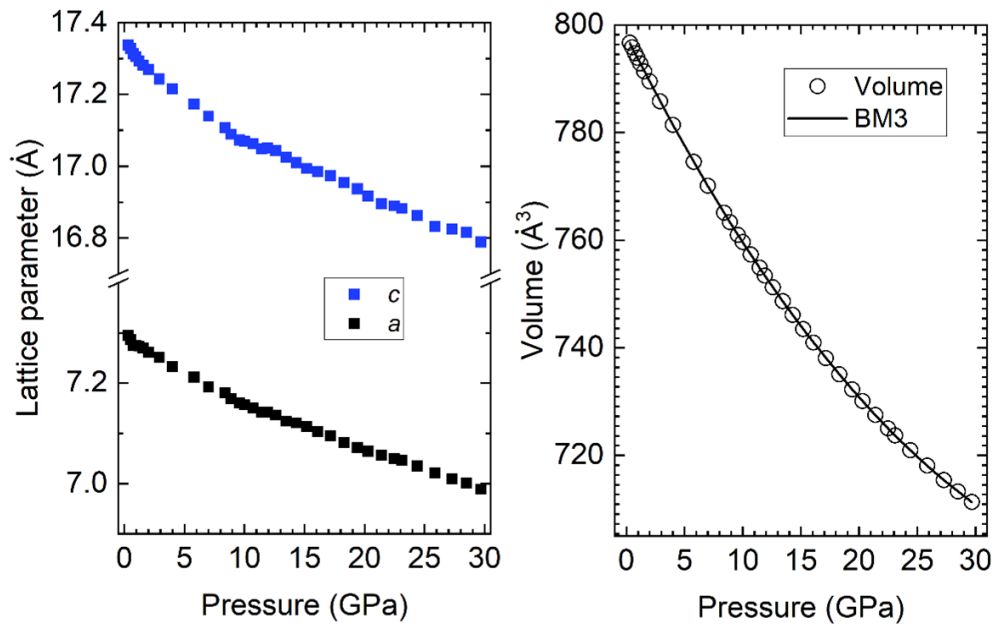

**Figure 16:** Pressure dependence of the unit-cell parameters and volume of Zirconolite-3T determined from Rietveld refinement of experimental XRD data. A third-order Birch-Murnaghan EOS has been fitted to the data and the fitted parameters are included.

## 4. Conclusions

In this work we reported a high-pressure synchrotron powder x-ray diffraction study of four different polytype structures of zirconolite, 2M, 4M, 3O, and 3T. In the case of zirconolite-2M, a structure different than that previously proposed was observed and has been named as zirconolite-2TR. We conclude that this structure is triclinic and can be found in the CCDC (deposition number 2246629). This finding is supported by density-functional theory calculations which showed that the previously reported structure [10] is dynamically unstable and has a higher enthalpy than the triclinic structure reported here. Both x-ray diffraction experiments and computer simulations support the existence in zirconolite-2TR of a phase transition at 14.7 GPa to a high-pressure monoclinic structure. This structure can be also found in the CCDC (deposition number 2268224). The existence of the phase transition in zirconolite-2TR agrees with the phase transition previously reported on zirconolite-2TR [20]. The comparison of both experiments suggests that the structural sequence of zirconolite-2TR could be $P\bar{1} \rightarrow C2/c(14.7\text{ GPa}) \rightarrow P2_1/m(\sim 40\text{ GPa})$. In the other three zirconolite structures studied in this work (4M, 3O, and 3T) there is no phase transition up to the highest pressure covered by experiments. We have also obtained information of the linear compressibility of the studied compounds as well as a room-temperature pressure-volume equation of state. The bulk moduli of the four compounds are similar with values close to 170 GPa. The compressibility seems to be driven by polyhedral compressibility of $CaO_8$ and $ZrO_7$ polyhedra. We consider the reported results are important for the design of nuclear-waste storage materials.

**Author Contributions**

Daniel Errandonea conceived the project. Lewis Blackburn and Neil C. Hyatt synthesized and characterized the samples. Robin Turnbull, Josu Sanchez-Martin, Robert Oliva, Jordi Ibañez-Insa, and Catalin Popescu performed high-pressure powder XRD experiments. Daniel Errandonea, Robin Turnbull, Robert Oliva, Jordi Ibañez-Insa performed data analysis. Alfonso Muñoz, Silvana Radescu, and Andres Mujica performed density-functional theory calculations. All authors participated in writing and editing of the manuscript. All authors have given approval to the final version of the manuscript.

**Declaration of competing interest**

The authors declare that they have no known competing financial interests or personal relationships that could have appeared to influence the work reported in this paper.

**Data Availability**

The data that support the findings of this study are available from the corresponding author upon reasonable request.

**Acknowledgements**

The authors gratefully acknowledge the financial support thank the financial support from the Spanish Ministerio de Ciencia e Innovación (DOI: 10.13039/501100011033) under Projects PID2019-106383GB-41/43, PID2022-138076NB-C41/44, and RED2022-134388-T. D.E. would also like to thank the financial support of Generalitat Valenciana through grants PROMETEO CIPROM/2021/075-GREENMAT and MFA/2022/007. This study forms part of the Advanced Materials program and is supported by MCIN with funding from European Union Next Generation EU (PRTR-C17.I1) and by the Generalitat Valenciana. The authors thank ALBA for providing beamtime under experiment No. 2022025734. R.T. acknowledges funding from the Generalitat Valenciana for Postdoctoral Fellowship No. CIAPOS/2021/20. J.S.-M. acknowledges the Spanish Ministry of Science, Innovation and Universities for the PRE2020-092198 fellowship. L. Blackburn wishes to acknowledge funding through the Royal Academy of Engineering Research Fellowship.

**Appendix A. Supplementary data**

The Supplementary data include information on the crystal structures studied in this work.

# Supplementary Material

# A comparative study of the high-pressure structural stability of zirconolite materials for nuclear waste immobilisation

Daniel Errandonea[1,*], Robin Turnbull[1], Josu Sánchez-Martín[1], Robert Oliva[2], Alfonso Muñoz[3], Silvana Radescu[3], Andres Mujica[3], Lewis Blackburn[4], Neil C. Hyatt[5,6], Catalin Popescu[7], Jordi Ibáñez-Insa[2]

[1]Departamento de Física Aplicada-ICMUV, MALTA-Consolider Team, Universidad de Valencia, Dr. Moliner 50, Burjassot, 46100 Valencia, Spain

[2]Geosciences Barcelona (GEO3BCN), MALTA-Consolider Team, Spanish Council for Scientific Research (CSIC), Lluís Solé i Sabarís s/n, 08028 Barcelona, Spain

[3]Departamento de Física, MALTA-Consolider Team, Instituto de Materiales y Nanotecnología, Universidad de La Laguna, San Cristóbal de La Laguna, E-38200 Tenerife, Spain

[4]Department of Materials Science and Engineering, The University of Sheffield, United Kingdom

[5]Building 329, Thomson Avenue, Harwell Campus, Didcot, OX11 0GD, United Kingdom

[6]School of Earth Sciences, The University of Bristol, Bristol, BS8 1RL, United Kingdom

[7]CELLS-ALBA Synchrotron Light Facility, Cerdanyola del Vallès, 08290 Barcelona, Spain

*Corresponding author: E-mail: daniel.errandonea@uv.es

Table S1: Crystal structure information of triclinic zirconolite-2TR at room temperature and 0.1 GPa.

| Composition: CaZrTi$_2$O$_7$ | Space group: $P\bar{1}$ | $a$ = 7.203(3) Å, $b$ = 7.216(3) Å, $c$ = 11.372(5) Å, $\alpha$ = 80.99(1)°, $\beta$ = 80.86(1)°, $\gamma$ = 60.59(1)° ($Z$ = 4) | | | |
|---|---|---|---|---|---|
| Atom | Wickoff position | x | y | z | Occupation |
| Ca1 | 2i | 0.24770 | 0.50065 | 0.00031 | 1 |
| Ca2 | 2i | 0.50007 | 0.75489 | 0.49010 | 1 |
| Zr1 | 2i | 0.00216 | 0.24271 | 0.52735 | 1 |
| Zr2 | 2i | 0.75587 | -0.00244 | -0.02543 | 1 |
| Ti1 | 2i | 0.14299 | 0.36515 | 0.75185 | 1 |
| Ti2 | 2i | 0.62838 | 0.87415 | 0.74601 | 1 |
| Ti3 | 2i | 0.87337 | 0.12985 | 0.25143 | 1 |
| Ti4 | 2i | 0.41790 | 0.52103 | 0.24688 | 1 |
| O1 | 2i | 0.20230 | 0.44446 | 0.22180 | 1 |
| O2 | 2i | 0.57825 | 0.83929 | 0.28229 | 1 |
| O3 | 2i | 0.33625 | 0.60705 | 0.39664 | 1 |
| O4 | 2i | 0.39001 | 0.67304 | 0.09797 | 1 |
| O5 | 2i | 0.11973 | 0.29643 | -0.07345 | 1 |
| O6 | 2i | 0.71043 | 0.8765 | 0.57040 | 1 |
| O7 | 2i | 0.05692 | 0.16417 | 0.70943 | 1 |
| O8 | 2i | 0.83993 | -0.07503 | 0.79192 | 1 |
| O9 | 2i | 0.25728 | 0.55388 | 0.77810 | 1 |
| O10 | 2i | 0.43257 | 0.77884 | 0.71361 | 1 |
| O11 | 2i | 0.53899 | 0.88172 | -0.08236 | 1 |
| O12 | 2i | 0.11633 | 0.46563 | 0.58400 | 1 |
| O13 | 2i | 0.87828 | 0.11775 | 0.08193 | 1 |
| O14 | 2i | 0.88012 | 0.12372 | 0.42087 | 1 |

Table S2: Crystal structure information of triclinic zirconolite-2TR at room temperature and 13.8 GPa.

| Composition: CaZrTi$_2$O$_7$ | Space group: $P\bar{1}$ | $a$ = 7.021(4) Å, $b$ = 7.033(4) Å, $c$ = 11.168(6) Å, $\alpha$ = 81.15(3)°, $\beta$ = 80.84(3)°, $\gamma$ = 60.55(3)° ($Z$ = 4) | | | |
|---|---|---|---|---|---|
| Atom | Wickoff position | x | y | z | Occupation |
| Ca1 | 2i | 0.2476 | 0.5005 | 0.0003 | 1 |
| Ca2 | 2i | 0.5002 | 0.7549 | 0.4902 | 1 |
| Zr1 | 2i | 0.0022 | 0.2428 | 0.5274 | 1 |
| Zr2 | 2i | 0.7559 | -0.0025 | -0.0254 | 1 |
| Ti1 | 2i | 0.1428 | 0.3652 | 0.7518 | 1 |
| Ti2 | 2i | 0.6284 | 0.8741 | 0.7460 | 1 |
| Ti3 | 2i | 0.8733 | 0.1299 | 0.2514 | 1 |
| Ti4 | 2i | 0.4179 | 0.5209 | 0.2468 | 1 |
| O1 | 2i | 0.2020 | 0.4443 | 0.2223 | 1 |
| O2 | 2i | 0.5772 | 0.8383 | 0.2824 | 1 |
| O3 | 2i | 0.3369 | 0.6072 | 0.3970 | 1 |
| O4 | 2i | 0.3898 | 0.6734 | 0.0983 | 1 |
| O5 | 2i | 0.1199 | 0.2958 | -0.0735 | 1 |
| O6 | 2i | 0.7106 | 0.8764 | 0.5699 | 1 |
| O7 | 2i | 0.0569 | 0.1633 | 0.7095 | 1 |
| O8 | 2i | 0.8391 | -0.0747 | 0.7921 | 1 |
| O9 | 2i | 0.2575 | 0.5537 | 0.7780 | 1 |
| O10 | 2i | 0.4329 | 0.7786 | 0.7136 | 1 |
| O11 | 2i | 0.5395 | 0.8826 | -0.0828 | 1 |
| O12 | 2i | 0.1171 | 0.4660 | 0.5837 | 1 |
| O13 | 2i | 0.8782 | 0.1177 | 0.0822 | 1 |
| O14 | 2i | 0.8801 | 0.1242 | 0.4211 | 1 |

Table S3: Crystal structure information of monoclinic high-pressure phase of zirconolite-2TR at room temperature and 14.7 GPa.

| Composition: CaZrTi$_2$O$_7$ | Space group: C2/c | $a$ = 12.035(9) Å, $b$ = 6.987(4) Å, $c$ = 11.304(9) Å, and $\beta$ = 100.12(2)° (Z = 8) | | | |
|---|---|---|---|---|---|
| Atom | Wickoff position | x | y | z | Occupation |
| Ca1 | 8f | 0.3732 | 0.1264 | 0.4958 | 1 |
| Zr1 | 8f | 0.1231 | 0.1221 | 0.9746 | 1 |
| Ti1 | 8f | 0.2495 | 0.1234 | 0.7467 | 1 |
| Ti2 | 4e | 0 | 0.5534 | 0.25 | 1 |
| Ti3 | 4e | 0 | 0.1264 | 0.25 | 1 |
| O1 | 2i | 0.3057 | 0.1268 | 0.2788 | 1 |
| O2 | 2i | 0.4706 | 0.1339 | 0.1006 | 1 |
| O3 | 2i | 0.2084 | 0.0852 | 0.5712 | 1 |
| O4 | 2i | 0.3987 | 0.1611 | 0.7206 | 1 |
| O5 | 2i | 0.7076 | 0.1732 | 0.5834 | 1 |
| O6 | 2i | 0.0033 | 0.1217 | 0.4189 | 1 |
| O7 | 2i | 0.1119 | 0.054 | 0.7905 | 1 |

Table S4: Crystal structure information of triclinic zirconolite-4M at room temperature and 0.1 GPa.

| Composition: $Ca_{0.75}Zr_{0.75}Dy_{0.50}Ti_2O_7$ | Space group: $C2/c$ | $a$ = 12.465(1) Å, $b$ = 7.189(1) Å, $c$ = 22.994(2) Å, $\beta$ = 84.82(1)° (Z = 16) | | | |
|---|---|---|---|---|---|
| Atom | Wyckoff position | x | y | z | Occupation |
| Ca1 | 4e | 0 | 0.1246 | 0.25 | 0.82 |
| Nd1 | 4e | 0 | 0.1246 | 0.25 | 0.18 |
| Ca2 | 8f | 0.7535 | 0.8732 | 0.2509 | 0.69 |
| Nd2 | 8f | 0.7535 | 0.8732 | 0.2509 | 0.3 |
| Ti1 | 4e | 0 | 0.6285 | 0.25 | 0.48 |
| Zr1 | 4e | 0 | 0.6285 | 0.25 | 0.52 |
| Ca3 | 8f | 0.8755 | 0.3773 | 0.4929 | 0.38 |
| Nd3 | 8f | 0.8755 | 0.3773 | 0.4939 | 0.61 |
| Zr2 | 8f | 0.8743 | 0.8756 | 0.4986 | 0.92 |
| Ti2 | 8f | 0.8743 | 0.8756 | 0.4986 | 0.08 |
| Ti3 | 8f | 0.0636 | 0.8839 | 0.3789 | 0.82 |
| Zr3 | 8f | 0.0636 | 0.8839 | 0.3789 | 0.18 |
| Ti4 | 8f | 0.7834 | 0.1360 | 0.3794 | 0.4 |
| Zr4 | 8f | 0.7834 | 0.1362 | 0.3794 | 0.1 |
| Ti5 | 8f | 0.4389 | 0.1155 | 0.6287 | 0.82 |
| Zr5 | 8f | 0.4389 | 0.1155 | 0.6287 | 0.18 |
| Ti6 | 8f | 0.1590 | 0.8640 | 0.6299 | 0.4 |
| Zr6 | 8f | 0.1590 | 0.8640 | 0.6299 | 0.1 |
| Ti7 | 8f | 0.8033 | 0.6243 | 0.3771 | 1 |
| O1 | 8f | 0.0075 | 0.1203 | 0.3623 | 1 |
| O2 | 8f | 0.8043 | 0.1133 | 0.4541 | 1 |
| O3 | 8f | 0.0617 | 0.9070 | 0.4598 | 1 |
| O4 | 8f | 0.9063 | 0.4250 | 0.3936 | 1 |
| O5 | 8f | 0.0632 | 0.8346 | 0.2895 | 1 |
| O6 | 8f | 0.7589 | 0.6158 | 0.4589 | 1 |
| O7 | 8f | 0.9092 | 0.8183 | 0.3884 | 1 |
| O8 | 8f | 0.3830 | 0.8786 | 0.6121 | 1 |
| O9 | 8f | 0.1785 | 0.8876 | 0.7043 | 1 |
| O10 | 8f | 0.4371 | 0.0927 | 0.7089 | 1 |
| O11 | 8f | 0.2818 | 0.5766 | 0.6425 | 1 |
| O12 | 8f | 0.4389 | 0.1657 | 0.5405 | 1 |
| O13 | 8f | 0.1340 | 0.3839 | 0.7082 | 1 |
| O14 | 8f | 0.2850 | 0.1827 | 0.6394 | 1 |

Table S5: Crystal structure information of triclinic zirconolite-3T at room temperature and 0.1 GPa.

| Composition: $Ca_{0.65}Ce_{0.35}ZrTi_{1.30}Fe_{0.70}O_7$ | | Space group: $P3_121$ | $a$ = 7.295(2) Å, $c$ = 17.337(3) Å ($Z$ = 6) | | |
|---|---|---|---|---|---|
| Atom | Wyckoff position | x | y | z | Occ. |
| Ca1 | 3a | 0.8355 | 0 | 0.3333 | 0.7 |
| Ce1 | 3a | 0.8355 | 0 | 0.3333 | 0.3 |
| Ca2 | 3a | 0.3309 | 0 | 0.3333 | 0.58 |
| Ce2 | 3a | 0.3309 | 0 | 0.3333 | 0.32 |
| Th1 | 3a | 0.3309 | 0 | 0.3333 | 0.1 |
| Zr1 | 6c | 0.1647 | 0.6682 | 0.0182 | 0.88 |
| Th2 | 6c | 0.1647 | 0.6682 | 0.0182 | 0.04 |
| Hf1 | 6c | 0.1647 | 0.6682 | 0.0182 | 0.02 |
| Zr2 | 6c | 0.1647 | 0.6682 | 0.0182 | 0.06 |
| Ti1 | 3b | 0.3289 | 0 | 0.8333 | 0.72 |
| Nb1 | 3b | 0.3289 | 0 | 0.8333 | 0.26 |
| Zr3 | 3b | 0.3289 | 0 | 0.8333 | 0.02 |
| Fe1 | 6c | 0.9011 | 0.0491 | 0.8327 | 0.36 |
| Mn1 | 6c | 0.9011 | 0.0491 | 0.8327 | 0.15 |
| Nb2 | 6c | 0.4985 | 0.3331 | 0.1637 | 0.47 |
| Ti2 | 6c | 0.4985 | 0.3331 | 0.1637 | 0.5 |
| Zr4 | 6c | 0.4985 | 0.3331 | 0.1637 | 0.03 |
| O1 | 6c | 0.5907 | 0.6236 | 0.1437 | 1 |
| O2 | 6c | 0.0052 | 0.8266 | 0.0591 | 1 |
| O3 | 6c | 0.5304 | 0.3075 | 0.0479 | 1 |
| O4 | 6c | 0.1971 | 0.2312 | 0.1443 | 1 |
| O5 | 6c | 0.5196 | 0.8912 | 0.0555 | 1 |
| O6 | 6c | 0.9441 | 0.3139 | 0.0552 | 1 |
| O7 | 6c | 0.2113 | 0.6249 | 0.1392 | 1 |

Table S6: Crystal structure information of triclinic zirconolite-30 at room temperature and 0.1 GPa.

| Composicion: $Ca_{0.20}Nd_{0.80}ZrTi_{1.20}Fe_{0.80}O_7$ | | | Space group: $Cmce$ | $a = 10.131(2)$ Å, $b = 14.049(2)$ Å, $c = 7.322(1)$ Å ($Z = 8$) | |
|---|---|---|---|---|---|
| Ca1 | 8e | 0.75 | 0.1159 | 0.25 | 0.529 |
| Na1 | 8e | 0.75 | 0.1159 | 0.25 | 0.045 |
| Th1 | 8e | 0.75 | 0.1159 | 0.25 | 0.017 |
| Ce1 | 8e | 0.75 | 0.1159 | 0.25 | 0.409 |
| Zr1 | 8f | 0.0155 | 0.2333 | 0.5 | 1 |
| Ti1 | 4a | 0 | 0 | 0 | 0.517 |
| Nb1 | 4a | 0 | 0 | 0 | 0.469 |
| Ta1 | 4a | 0 | 0 | 0 | 0.014 |
| Ti2 | 8e | 0.25 | 0.1339 | 0.25 | 0.878 |
| Nb2 | 8e | 0.25 | 0.1339 | 0.25 | 0.119 |
| Ta2 | 8e | 0.25 | 0.1339 | 0.25 | 0.003 |
| Fe1 | 8d | 0 | 0 | 0.4306 | 0.462 |
| Fe2 | 8f | 0.0396 | 0.0157 | 0.5 | 0.029 |
| O1 | 16g | 0.1263 | 0.0333 | 0.1913 | 1 |
| O2 | 16g | 0.1197 | 0.2331 | 0.2104 | 1 |
| O3 | 8f | -0.1013 | 0.1093 | 0.5 | 1 |
| O4 | 8f | -0.0914 | 0.1299 | 0 | 1 |
| O5 | 8f | 0.1803 | 0.1399 | 0.5 | 1 |